\documentclass[11pt]{article}
\usepackage{moriond,epsfig}

\def\be{\begin{equation}}
\def\ee{\end{equation}}
\def\bea{\begin{eqnarray}}
\def\eea{\end{eqnarray}}

\begin{document}
\vspace*{4cm}
\title{Recent Results from the AMANDA-II neutrino telescope}

\author{ Andreas Gro{\ss} for the AMANDA collaboration }

\address{Institute of Physics, University of Dortmund, 44221 Dortmund}


\maketitle\abstracts{
AMANDA-II is an operating neutrino telescope located at the South Pole.
Recent results of AMANDA are presented, including the examination of the 
diffuse neutrino flux, permanent and transient point source analyses,
and indirect dark matter searches. A brief outlook on the IceCube neutrino
telescope currently under construction at the South Pole is
given.
}

\section{Introduction}
The observation of high-energy cosmic rays - both charged particles and
photons - motivates the
search for extraterrestrial high-energy neutrinos.
Charged particles have been detected up to energies around $10^{20}$ eV while
photons up to tens of TeV could be attributed to point sources. Most of
the 
production mechanisms of highly energetic nuclei as well as hadronic
models for the TeV 
$\gamma$ radiation imply the coproduction of high-energy neutrinos.
This allows neutrinos to be used as astrophysical messengers giving additional
information on the production mechanisms in different astrophysical
environments. 
Due to their small cross section and their lack of an electrical charge,
$\nu$'s do neither suffer from deflection nor absorption on the way from the
source to the Earth. 
This small cross section, however, requires very large detector
volumes in order to provide the observer with sufficiently high $\nu$ induced
event rates. 
\section{The AMANDA neutrino telescope}
The AMANDA neutrino telescope at the geographic south pole was built in the
years 1995-2000 in the antarctic ice sheet. 
AMANDA-II consists of 677 optical modules (OMs) containing
photomultiplier tubes (PMTs) on 19 strings in the south pole ice\footnote{The
  AMANDA-B10 subdetector was running from 1997-1999 with 302 OMs on the inner
  10 strings.}. The 
geometrical shape of the detector is a  cylinder with a diameter of
$200$ m. Most modules are concentrated at depths between $1500$ m-$2000$ m. 

Most of the events triggered by AMANDA are atmospheric muons reaching the
detector from above. 
Muons produced in charged current (CC)
interactions of $\nu_\mu$ can be separated from atmospheric ones by taking
into account directional and energy information. 

The Earth serves as a shield against any other particles, such that 
upgoing muons must be $\nu_\mu$ induced.  The lower AMANDA threshold for
$\nu_\mu$-induced events is around 50 GeV, while at PeV energies, the
Earth becomes opaque for $\nu_\mu$. Thus, for energies below a few PeV,
AMANDA's field of view for 
$\nu_\mu$ is restricted to the northern sky. At higher energies, the search for
downgoing 
$\nu_\mu$ induced events is possible since the background of atmospherical
muons falls steeply with energy.
Muons travel long distances through the ice emitting continously Cherenkov
light. This increases the effective volume of the detector and provides a
long lever arm 
for track reconstruction. An angular resolution of $3^\circ$ and
below has been achieved, depending on the declination angle and on the
strength of quality cuts.  
Other $\nu$ flavors can be detected by the light produced by hadronic- and
electromagnetic cascades, if the interaction takes place
within the detector. In the cascade channel, energy resolution is
much better, but the event rates are lower. Due to the lack of a long lever
arm, the angular resolution is
only about $30^\circ$. Here, we focus on the $\nu_\mu$
channel and energies below $1$~PeV. 

\section{Cosmic accelerators}
The existence of highly energetic cosmic rays can be explained by the mechanism
of Fermi acceleration. Statistically, a charged particle gains energy passing
an 
electromagnetic shock front. The random energy gain results in a power
law spectrum for the cosmic ray flux with a spectral index $\gamma \approx
2$. A review on cosmic ray production by shock acceleration can be found
in~\cite{prtheroe_SI}. 

The accelerated protons interact with each other or ambient photons. Thus,
charged and neutral pions are produced mostly via the $\Delta$ resonance. Pion
decay yields a $\nu$ flux with roughly the same spectral index as the
protons. At production site, the $\nu$ flavors are distributed according to
$\nu_e:\nu_\mu:\nu_\tau \approx 1:2:0$. Due to $\nu$ oscillations, an
approximate equipartition of flavors is predicted to be observed at the
detector.

The extremely high total energy needed for particle acceleration to ultra-high
energies (UHE) 
energies suggests the gravitational potential of a black hole or a neutron
star as the source.  
Likely astrophysical environments for shock acceleration are
jet-disc systems like Active Galactic Nuclei (AGN), Gamma Ray Bursts
(GRBs) and Microquasars as well as shock fronts from Supernova explosions
expanding into the interstellar medium or into the wind of the preceding star.

Additional to the described shock acceleration mechanism, top-down scenarios
for the production of cosmic rays are discussed. These models assume the decay
of heavy non-Standard Model (SM) particles (see e.g. \cite{drees}).
\section{Physics analysis strategies}
After selecting upgoing muon tracks with high quality reconstruction fits, 
the resulting AMANDA event samples are strongly dominated by atmospheric
$\nu_\mu$. Several analysis strategies have been developed to analyze these
atmospheric $\nu_\mu$ and to search for astrophysical $\nu_\mu$.
In the diffuse analysis, extraterrestrial $\nu_\mu$ can be detected by a
significant excess of the number of highly energetic $\nu_\mu$ over the
expectation from the atmospheric $\nu_\mu$ flux.
In point source analyses, the directional information is used to search for an
excess of  $\nu_\mu$ coming from the direction of a potential source. For
time-dependent sources like GRBs, the arrival time information can be used for
a further background reduction. 

The AMANDA collaboration follows a strict blindness policy, i.e.\
an analysis has to be developed blindly with respect to the data. That
ensures a statistical correct analysis.
In practice, that means that all optimizations are done on a subsample of data
that is not
used for the final result or that the relevant quantities are randomized
(e.g.\ reconstructed right ascension  in a point source analysis).
\section{Diffuse neutrino spectrum}
The energy losses of muons in the ice consist of an almost energy independent
part by due to ionization and energy-dependent stochastic losses due to
radiative processes. The energy loss  is parameterized as dE/dx = a + bE,
where $a/b \approx 600$ GeV. Thus the amount of Cherenkov light deposited
turns out to 
be proportional to the energy and allows for energy reconstruction.

With help of a  full Monte Carlo simulation chain, some energy dependent
variables were identified, e.g.\ the number of hit OMs. 
A neural net was trained with a high number of MC events using $6$ input
 variables and detector settings of the year 2000. For each measured event an 
energy correlated parameter is determined 
by the neural net.
In order to correct for the limited acceptance and for the finite resolution of
the detector, the method of regularized unfolding is used \cite{blobel}. The
correction for 
the energy losses at the vertex is also done within the unfolding.  
With the input of
the energy correlated parameter from the neural net and two
further energy dependent variables for all events, the measured $\nu_\mu$
spectrum is determined.

\begin{figure}[htb]
\epsfig{file=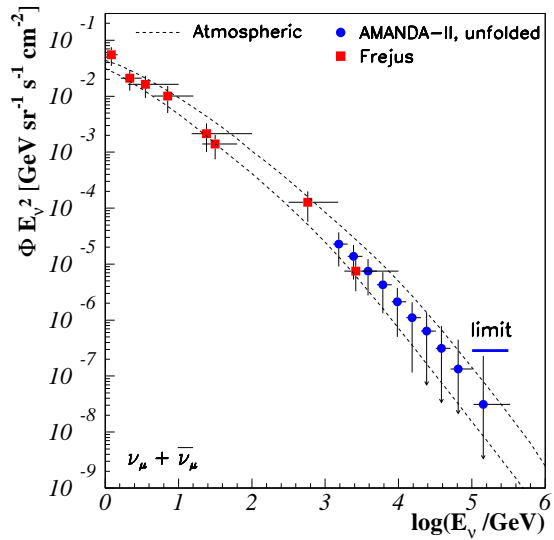,width=8cm,angle=0}
\caption{Unfolded neutrino ($\nu_\mu +\bar{\nu }_\mu$) energy
  spectrum as measured by AMANDA~II and Frejus. The AMANDA~II
  limit on the $E^{-2}$ flux is marked by the solid line above the highest
  energy bin.}
\label{unfolded spectrum}
\end{figure}

The preliminary unfolded $\nu_\mu$ spectrum of AMANDA~II based on  events from
the year 
2000 is displayed in Fig.\ \ref{unfolded spectrum} together with lower energy 
measurements of Frejus \cite{frejus_spectrum}. All data points are in  
agreement with the expectation for the atmospheric $\nu_\mu$ flux. For
energies above $100$ GeV, the
expectation is taken from the prediction in \cite{Volkova}, for lower energies
the parametrization from \cite{honda} is used. The prediction for the
horizontal flux is given by the upper line, while the lower line corresponds
to the expectation for the vertical flux. 

The unfolded spectrum provides one possibility to calculate a
limit on an additional $E^{-2}$ flux. The highest energy bin at
$100$~TeV-$300$~TeV was determined to be most sensitive to the additional 
contribution. 
Since the statistics of the unfolding output is unknown a priori, the
probability density function (pdf) of the results from the unfolding program
has to be determined with  Monte Carlo simulations. For various signal
contributions between $10^{-8} 
\mbox{GeV} \mbox{s}^{-1} \mbox{cm}^{-2} \mbox{sr}^{-1}$ and $10^{-6} 
\mbox{GeV}  \mbox{s}^{-1} \mbox{cm}^{-2} \mbox{sr}^{-1}$, large numbers of MC
experiments are performed.  
The pdf is extracted from histograms of the output of the unfolding program
for the MC experiments.
With the pdf, Feldman-Cousins confidence belts \cite{FC} are constructed. 
By using these confidence belts, the $90\%$~C.L.\ upper limit for the year 2000
data is determined preliminarily   
to $$E^2\Phi < 2.6\cdot 10^{-7} 
\mbox{GeV} \mbox{s}^{-1} \mbox{cm}^{-2} \mbox{sr}^{-1}\,,$$including $33\%$ for
systematical uncertainties 
(mainly due to unknown ice parameters).

\section{Point source analysis}
In addition to diffuse analyses, the directional information of $\nu_\mu$
events can be used in a point source analysis to identify individual $\nu$
sources. 
From the data collected in the years 2000-2003 (livetime 807 days), an event
sample was extracted for point source analyses  containing
3329 upgoing tracks. Quality cuts on the fitted tracks ensure a good
angular resolution and a minimal 
contamination by misreconstructed atmosperic muons. The event selection was
optimized for best sensitivity to point sources with spectral indices $\gamma$
between
$\gamma=-3$ and $\gamma=-2$.
The event sample was searched for a significant excess of $\nu_\mu$ from a
point source.
\subsection{Steady point sources}
Since $\nu$ production via pion decay implies the co-production of high
energetic $\gamma$ radiation, strong $\gamma$ sources are good candidates for
highly energetic $\nu$ emission.
Thus, a list of $33$ pre-selected source candidates, mainly
consisting of high energy $\gamma$ sources, was analyzed.
The candidate list covers  $13$ blazars, $3$ nearby AGN and galactic source
candidates such as microquasars and  supernova remnants. 
For none of the $33$ selected sources, a significant signal was found. The
highest excess was found for the Crab nebula with 
$10$ events at a background of $5.36$ events. The background probability is
given by $\log{P} = -1.35$. Considering the trial factors, such an
excess is
expected for one of the 33 sources with $64\%$ probability in a
pure background sample.

Additionally, in an all-sky search, $\nu$ point sources at any point of the
sky have been searched the by a grid of overlapping bins. These bins were
tested for a significant 
excess over the background expectation determined by the number of events
coming from the same zenith band. The highest significance in any bin (before
correction for the trial factor) was found with $3.35 \sigma$. A toy
MC shows that this value can be expected with a probability of $92\%$ in a
random sample. 

\begin{figure}[htb]
\epsfig{file=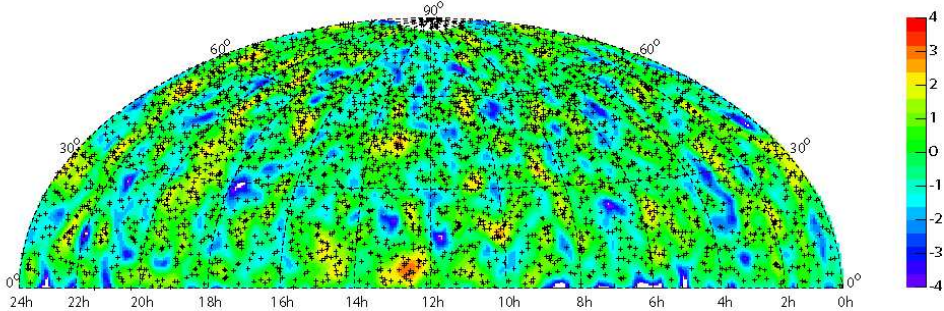,width=13cm,angle=0}
\caption{Significance sky map of the point source analysis (preliminary). The
  dots correspond to the reconstructed arrival directions of individual
  events. }
\label{skymap}
\end{figure}

Furthermore, an unbinned method was used, calculating the significance for an
excess over the background given the track direction and the resolution
of the individual reconstructed events, as determined from the fit
uncertainties \cite{till}. The significance sky map from this method, as
displayed in Fig.\ \ref{skymap}, 
shows no deviation from a sky map of the same sample with randomized right
ascension.

Hence, no statistically significant excess was found from any steady point
source. On all source candidates, upper limits were calculated. 
The preliminary results presented here update the published results for data
collected in 2000-2002 \cite{paolo}.
\subsection{Source stacking of AGN}
In order to increase the sensitivity of telescopes, so called source stacking
methods have been applied in $\gamma$ astronomy and in optical astronomy. In a
source stacking analysis, the cumulative signal of several sources of the same
type is evaluated. A systematical classification is required to define the
relevant source samples.

For AMANDA, a source stacking analysis of AGN has been developed based on a
phenomenological classification. The axisymmetry of AGN is one central ordering
parameter yielding different appearance of AGN depending on the observer's
angle to the AGN axis. 
If this angle is small, an AGN appears as a blazar and emission from the jet
gets boosted. 
A further difference is the strength of the radio flux
of the AGN, resulting in a classification into radio-loud AGN and radio-weak
ones.
 
The $\nu$ production mechanism via $\Delta$ resonance and pion decay implies
correlations between the photon spectrum and the $\nu$ spectrum at the
production site. 
These correlations may be distorted, if sources are optically thick for
photons. 
In case of optically thick sources, the high energy
photons  induce electromagnetic cascades. Then  photon energies get
reprocessed and the photon flux is shifted to lower energies.
Taking into account the high cross sections of photons, the current knowledge
about AGN does not exclude optically thickness effects 
for high energy photons in AGN.

The possibility of optical thickness motivates a selection also at lower
$\gamma$ 
energies than the AMANDA threshold for $\nu$. In total, 11 samples were
compiled.  
The number of sources to be included in the source
stacking analysis is optimized with the hypothesis that the relative TeV $\nu$
flux of sources is proportional to the relative photon flux at selection
energy. For details of the source selection see \cite{stacking}.

An application of the stacking method with the year 2000 data on these source
classes showed no significant excess over the background expectation.
We calculated $90\%$ C.L. upper limits on the integral flux of $\nu_\mu$ above
10 GeV of $(0.1-1.3)\cdot 10^{-8} \mbox{cm}^{-2} \mbox{s}^{-1}$ per source.
This is significantly 
smaller than the sensitivity \footnote{ average $90\%$ C.L. upper limit on
  flux in case of absence of signal} of $2\cdot 10^{-8} \mbox{cm}^{-2}
\mbox{s}^{-1}$ of the 
point source analysis for individual sources of that sample.


\subsection{Time-variable studies of preselected sources}
Additionally to the searches for steady point sources, a search for a
time-variable signal has been done. In this search, several
candidate sources which are known to be highly variable were analyzed. Two
different search 
strategies have been used:
\begin{itemize}
\item \emph{Search for clusters of events in coincidence with identified
  periods of enhanced photon emission - Multiwavelength approach:}
  Three objects have been analyzed, two blazars (Mkn~421 and 1ES~1959+650) and
  a microquasar (Cyg X-3). The active periods have been 
  defined by the X-ray light curves for the blazars and by the radio light
  curve for the microquasar.
\item \emph{Search for occasional $\nu$ bursts:}
A catalogue of $12$ highly variable source candidates was investigated. No
assumption on the time of the occurrence of $\nu$ flares is made a priori. A
sliding time window of fixed duration is searched for excesses of events. The
duration of the time window is set to $20$ days for galactic sources and to
$40$ days for extragalactic ones. Monte Carlo simulations of various scenarios
of flare duration support this choice of the duration.
\end{itemize}
The searches have been performed following the AMANDA blindness principle such
that biases which cannot be quantified are avoided. As a preliminary result, in
both approaches, no signi\-fi\-cant excess has been found for any of the
sources 
analyzed. In the multiwavelength approach, the highest excess was found for 1ES
1959+650 with 2 events on a background of 1.57 events. In the search for
occasional $\nu$ bursts, for GRS 1915+105 and for 1ES 1959+650 doublets have
been found with background probabilities of $32\%$ and $34\%$, respectively.

In an a-posteriori analysis of the 1ES 1959+650, the arrival times of the
individual $\nu$'s were considered. One of the events is within a few hours
from a reported flux peak in the 
Whipple light curve \cite{whipple1es}. Since X-ray measurements do not show a
peak, this peak is  considered as an "orphan flare".
An "orphan flare" is a flare in the high
energy gamma-ray range (TeV) without a coincident increase in X-ray (keV)
emission. As active periods are selected on the basis of the X-ray light
curves,
this event is not present in the preselected active periods. Exceptions from
the
usually observed correlations between keV and TeV fluxes are seen as an
indication of hadronic mechanisms responsible for TeV emission.

Even though the theoretical background draws attention to this observation, it
must be stressed that the probability of a random coincidence of the AMANDA
event with the time of the flare is difficult to evaluate. Additionally, in all
reasonable scenarios, a signi\-fi\-cant detection at a $5\sigma$ level can be
excluded. 

\section{Gamma Ray Bursts}
Gamma Ray Bursts (GRBs) are transient phenomena emitting a very high 
amount of $\gamma$ radiation in a very short time.
Under the assumption of isotropic radiation, their luminosity reaches 
$10^{51}$ erg/s. Typical burst durations vary between 0.1 s and 1000 s.
While their nature remained unknown for a long time, the fireball model 
recently turned out to describe the phenomenology successfully.
According to the fireball picture, GRBs are formed by the collapse of a massive
precessor object. Similar to AGN, the collapse is not rotational symmetric
and the conservation of angular momentum results in the occurrence of two jets
along the rotation axis.
In these jets, shock waves propagate and when reaching the surface, 
they emit the observed $\gamma$ radiation. The beaming 
effects reduce the necessary energy supply of the GRB. A connection of GRBs to
supernovae  of type Ic is confirmed by observations of GRB030329 and
GRB/980425. 
The diffuse energy spectrum of $\nu$ from GRBs has been predicted by the 
assumption that $\nu$ and $\gamma$ are produced by the same hadronic mechanism
\cite{WB}. This results in the broken power law spectrum, which we will refere
to as the Waxman-Bahcall (WB) spectrum. 

The location in space \emph{and} in time allows to search with AMANDA for $\nu$
from GRBs at a very low background.  An analysis of $139$ bursts in the years
2000-2003 was optimized for the WB spectrum. The time interval was
chosen to cover $90\%$ of the photon signal. For none of the bursts, an event
was found in that time window. The total background expectation is
$1.25$ events. From that 
result, a preliminary $90\%$ C.L.\ upper limit on the WB spectrum of 
$E^2\Phi < 3 \cdot 10^{-8} \mbox{GeV  cm}^{-2}\mbox{s}^{-1}$ was calculated.  

The possibility of $\nu$ leaving the source while it still remains optically
thick for $\gamma$ is taken into account by a precessor analysis, covering the
time before the measured burst. Again, no signal was found. In the future, an
optimization of the analysis for individual burst parameters can increase the
sensitivity.

\section{Indirect WIMP searches}
The minimal supersymmetric extension of the standard model (MSSM)
provides a promising dark matter candidate, the lightest neutralino 
$\tilde\chi_0^1$,  a linear combination of the supersymmetric partners of the
electroweak neutral gauge and Higgs bosons. Assuming R-parity conservation, the
$\tilde\chi_0^1$ is stable. 
according to this scenario, trapping in the gravitational
potential of the Sun or the Earth may increase the neutralino density
there. Then, pairwise annihilation gets enhanced and
the decay of the annihilation products yields a high energy
neutrino flux. 
A high energy neutrino flux from these regions can only be interpreted as the
annihilation of non-SM particles, since the solar
neutrino flux from the standard model is restricted to MeV energies.

\begin{figure}[htb]
\epsfig{file=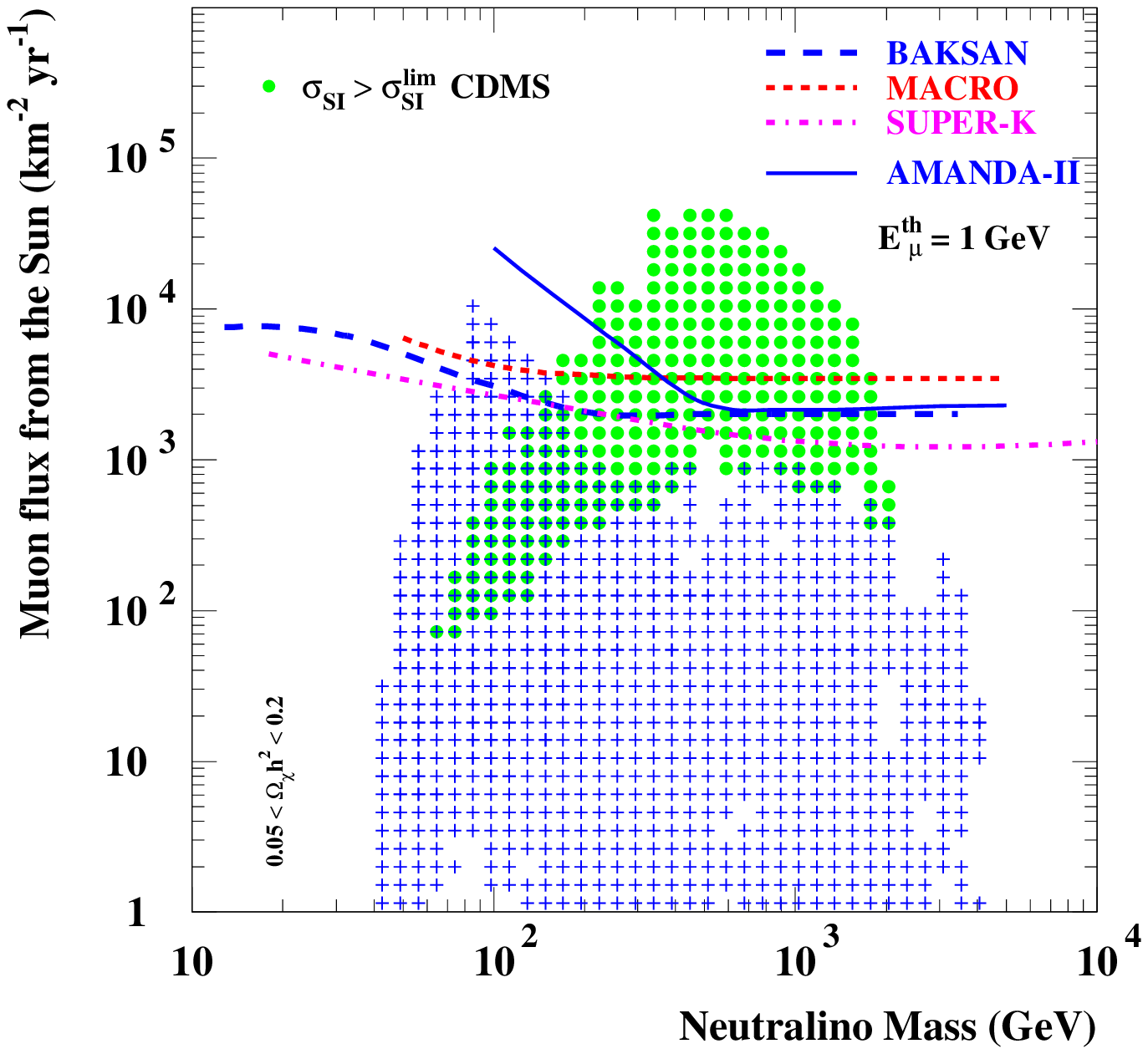,height=7cm,angle=0}
\epsfig{file=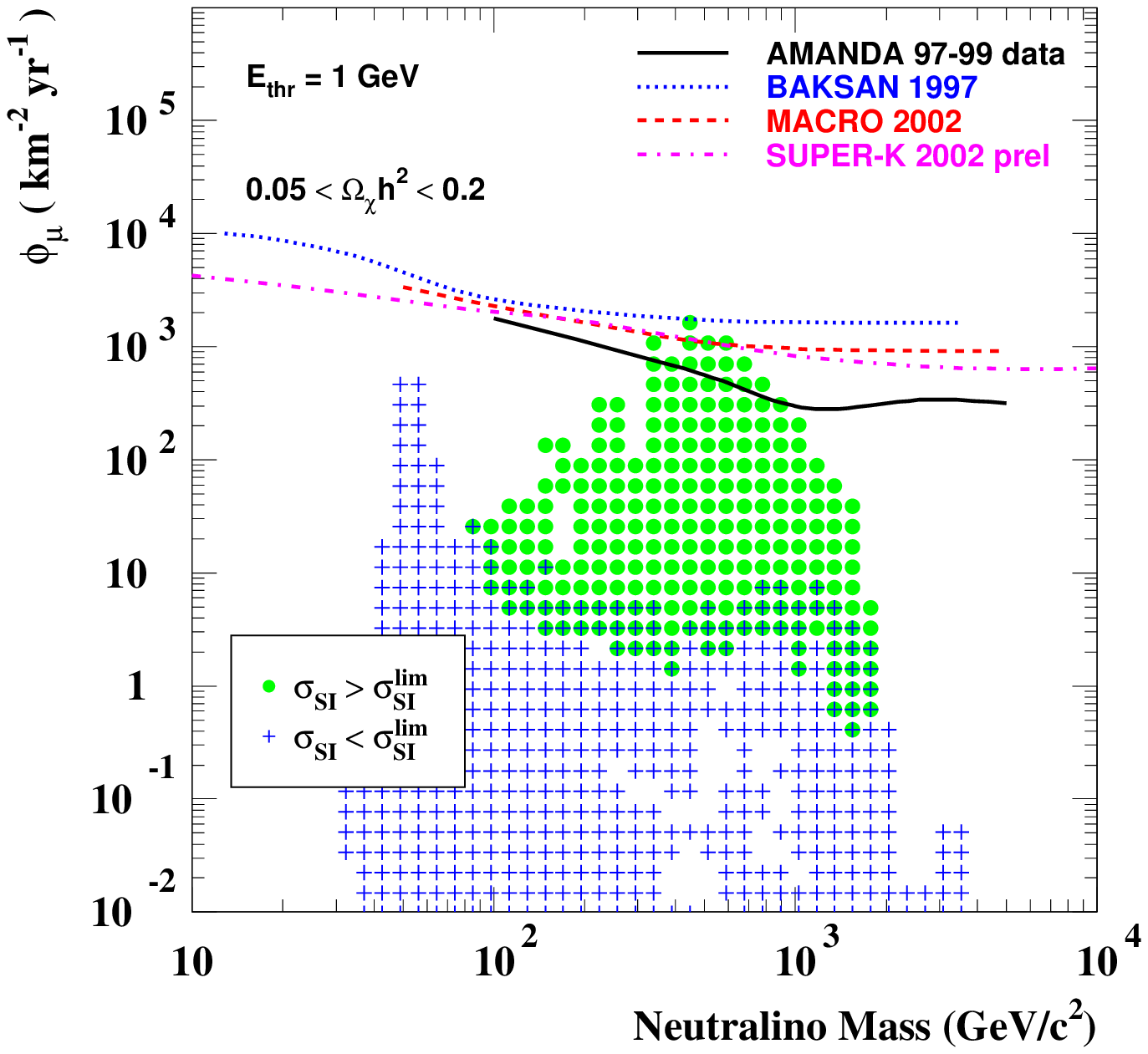,height=7cm,angle=0}
\caption{Limits on muon flux from WIMPs annihilating in  the Sun (left) or in
  the Earth (right). }

\label{WIMPs}
\end{figure}
The sensitivity of AMANDA  towards SUSY dark matter has been investigated by
Monte Carlo simulations with the program DARKSUSY \cite{darksusy}.
For different neutralino masses between $50$ GeV and $5$ TeV, two annihilation
channels were considered: The soft channel
$\tilde\chi_0^1\tilde\chi_0^1\rightarrow b \bar{b}$ and the hard channel
$\tilde\chi_0^1\tilde\chi_0^1\rightarrow W^+ W^-$. The terms soft and hard
refer to the fraction of total energy being emitted in $\nu$.

The size of the search bin was optimized for the expected neutrino spectrum
for the different channels and neutralino masses.
For solar WIMPS and data collected in 2001, the optimal bin size varies
between $5.5^\circ$ for the hardest spectrum and $26^\circ$ for the softest
spectrum. In this analysis, no significant excess over the atmospheric
background was found. The resulting limits  on the integrated muon flux 
for the hard channel are
plotted in Fig.\ \ref{WIMPs} on the left hand side. Limits from other
indirect searches 
e.g.\ from SUPER-K are also plotted. In that plot, the points represent the
expectation for different SUSY parameters. The expectations are calculated
with DARKSUSY. Values disfavored by the
direct search for WIMPs in CDMS II \cite{CDMS}
marked by darkerpoints. The restrictions on the parameter space due to the
direct search appear more severe. However, it should be noted the two methods
are complementary. They probe the WIMP distribution in the solar system at
different epochs and they are sensitive to different parts of the velocity
distribution. A detailed discussion of the results of the solar WIMP analysis
with AMANDA can be found in \cite{solarWIMPs}.

For WIMPs from the Earth, data collected with the AMANDA-B10 subdetector in
 1997-1999  (livetime 433 days) was analyzed with the same method. No
 significant excess 
 over the background expectation was found. In the same way as for solar
 WIMPs, the resulting preliminary limits of that 
analysis are displayed in Fig.\ \ref{WIMPs} on the right hand side.

\section{Conclusions and outlook}
The AMANDA-II neutrino telescope has been operating successfully since 2000 and
yielded remarkable results.
AMANDA has measured the atmospheric $\nu_\mu$ flux up to $100$ TeV. No
extraterrestrial $\nu$ flux has been detected so far. The upper limits on the
 flux, as set in the
different analyses, already constrain various theoretical models. 

While the analysis of AMANDA data goes on, 
in January 2005, the construction of the IceCube neutrino telescope
\cite{IceCube} began with
the successful deployment of the first string with 60 digital OMs
(DOMs). In the final stage, IceCube will reach a cubic kilometer size
instrumented with 4800 DOMs on 80 strings. The surface array IceTop will have 
stations near the top of each string. Each station will consist of 2 tanks of
frozen water viewed by standard IceCube DOMs. IceTop
 will allow for coincident surface and in ice
measurements of air showers induced by cosmic rays. 

The existing AMANDA strings will be
integrated in IceCube as a more densely instrumented detector region. The
increase in sensitivity with IceCube is 
considered to be sufficient for the detection of astrophysical $\nu$ fluxes in
most theoretical models. 

\section*{Acknowledgments}
{\small We acknowledge the support from the following agencies:
 The U.S. National Science Foundation,
the University of Wisconsin Alumni Research Foundation,
the U.S. DoE,
the U.S. NERS Computing Center,
the UCI AENEAS Supercomputer Facility,
the Swedish Research Council,
the Swedish Polar Research Secretariat,
the Knut and Alice Wallenberg Foundation (Sweden),
the German Federal Ministry of Education and Research,
the Deutsche Forschungsgemeinschaft,
the IWT (Belgium),
the FWO(Belgium),
the FNRS (Belgium)
and the OSTC (Belgium).
 D.~F.~Cowen acknowledges the support of the NSF CAREER
 program.
}

\end{document}